# TETRA Observation of Gamma Rays at Ground Level Associated with Nearby Thunderstorms

Rebecca Ringuette, Gary L. Case†, Michael L. Cherry, Douglas Granger,
T. Gregory Guzik, Michael Stewart, John P. Wefel

Dept. of Physics and Astronomy, Louisiana State University, Baton Rouge, LA 70803
† Currently at Department of Physics, La Sierra University, Riverside, CA 92515

Corresponding author: M.L. Cherry, Dept. of Physics and Astronomy, Louisiana State University, Baton Rouge, LA 70803 (cherry@lsu.edu)

**Abstract:** Terrestrial Gamma ray Flashes (TGFs) -- very short, intense bursts of electrons, positrons, and energetic photons originating from terrestrial thunderstorms -- have been detected with satellite instruments. TETRA, an array of NaI(Tl) scintillators at Louisiana State University, has now been used to detect similar bursts of 50 keV to over 2 MeV gamma rays at ground level. After 2.6 years of observation, twenty-four events with durations 0.02- 4.2 msec have been detected associated with nearby lightning, three of them coincident events observed by detectors separated by ~1000 m. Nine of the events occurred within 6 msec and 3 miles of negative polarity cloud-to-ground lightning strokes with measured currents in excess of 20 kA. The events reported here constitute the first catalog of TGFs observed at ground level in close proximity to the acceleration site.

**1. Introduction:** Lightning provides the most powerful natural accelerator available on Earth for producing high energy particles. Intense millisecond-scale bursts of gamma rays produced by upward-moving electrons accelerated to energies of tens of MeV or more have been detected with satellite instruments. These Terrestrial Gamma Flashes (TGFs) have been shown to be associated mainly with positive polarity intracloud lightning, with the particle acceleration occurring at altitudes of 10-15 km. We show here that negative polarity cloud-to-ground lightning accelerates particles downward and produces gamma rays with energies of at least 2 MeV. We present a sample of 24 TGFs detected at ground level associated with nearby (< 3 miles) lightning observed over approximately 2.6 years mainly during spring and summer thunderstorms in Louisiana.

TGFs were first observed by the Burst and Transient Source Experiment (BATSE) aboard the Compton Gamma Ray Observatory [Fishman et al., 1994; Gjesteland et al., 2012] and have now been observed by five additional satellite detectors – RHESSI [Smith et al., 2005; Grefenstette et al., 2009], the Gamma-Ray Imaging Detector (GRID) [Marisaldi et al., 2010] and MiniCALorimeter (MCAL) [Marisaldi et al., 2011; Tavani et al., 2011] on AGILE, and the Gamma ray Burst Monitor (GBM) [Cohen et al., 2010; Fishman et al., 2011; Briggs et al., 2013] and Large Area Telescope (LAT) [Grove et al., 2012] on the Fermi mission. Events are typically detected close to the sub-satellite point [Grefenstette et al., 2009] and are correlated both with regions of high thunderstorm activity [Cohen et al., 2006; Fuschino et al., 2011; Marisaldi et al., 2011] and with individual positive polarity intracloud (+IC) and possibly positive cloud-to-ground (+CG) lightning discharges to within 1 - 2 msec [Inan et al., 2006; Stanley et al., 2006; Hazelton et al., 2009]. (Positive polarity is needed to produce the upward beam of electrons and secondary photons necessary for detection of TGFs from space [Dwyer, 2003; Cohen et al., 2010].) Lightning flashes are known to emit a large fraction of their electromagnetic energy into low frequency (0.3 – 30 kHz) atmospheric radio signals (sferics), which can be located accurately by arrival time measurements in a worldwide radio receiver network [Rodger et al., 2009]. TGFs are well correlated both with sferics [Inan et al., 2006; Connaughton et al., 2013] and the LIS-OTD and WWLLN high resolution lightning data [Hazelton et al., 2009; Smith et al., 2010; Fuschino et al., 2011].

GBM has also demonstrated that in some cases, as the original gammas propagate upward through the atmosphere, they produce secondary $e^{\pm}$ via pair production [Cohen et al., 2010; Briggs et al., 2011] that escape into space. These secondaries are then able to spiral around magnetic field lines to the spacecraft far from the lightning location, producing Terrestrial Electron Beams (TEBs) characterized by 511 keV signals and both long duration pulses and delayed pulses resulting from particles moving past the spacecraft and then reflecting from magnetic mirror points and returning to be detected by GBM.

Given the altitude of the satellites around 500 km, the observations point to beaming of the photons upward with a ~30º half-angle cone coupled with attenuation of wide-angle photons passing through greater atmospheric path lengths [Grefenstette et al., 2008;

Østgaard et al., 2008; Hazleton et al., 2009; Gjesteland et al., 2011]. Based on the spectra observed by RHESSI [Smith et al., 2005], Dwyer and Smith [2005] performed detailed Monte Carlo simulations showing that the spectra were consistent with bremsstrahlung from electrons accelerated by the relativistic runaway electron avalanche (RREA) mechanism [Gurevich et al., 1992; Dwyer, 2003] at altitudes near thunderstorm tops. Over the 0.1 - 10 MeV range, the spectrum observed by AGILE [Marisaldi et al., 2011] has been well fit by a cutoff power law of the form $F(E) \sim E^{-\alpha} e^{-E/E_o}$ with $E_o$ compatible with the ~7 MeV electron energies predicted by RREA, but the observation of individual gamma rays with energies in excess of 40 MeV has posed a challenge for the emission models [Tavani et al., 2011; Celestin et al., 2012].

RREA models of TGFs in the atmosphere [Dwyer, 2003, 2008; Gjesteland et al., 2011] start with MeV seed electrons accelerated when lightning-associated electric fields overcome local energy losses. These accelerated electrons produce photons, secondary electrons, positrons, and X-rays by bremsstrahlung, pair creation, and Compton scattering. The avalanche may be further seeded by the relativistic feedback mechanism, in which backward propagating positrons and X-rays lengthen the TGF durations up to several milliseconds as seen in satellite observations [Dwyer, 2008; Dwyer et al., 2012a and references therein]. Ground-based lightning observations and comparisons of model calculations with the measured spectra indicate that the TGFs are produced at altitudes ~ 10 - 25 km  [Dwyer and Smith, 2005; Grefenstette et al., 2008; Shao et al., 2010; Gjesteland et al., 2010; Cummer et al., 2011; Xu et al., 2012]. A more detailed review of TGF models and observations is presented by Dwyer et al. [2012b].

TGF observations from satellite platforms are limited to events apparently beamed upward and large enough to be detected even in the presence of attenuation and Compton scattering by the atmosphere. Although these events observed from space are extremely intense (gamma ray rates in excess of 300 kHz measured with BATSE), the bulk of the events are presumably smaller events which can only be observed much closer to the lightning -- i.e., at aircraft or balloon altitude or at ground level [Smith et al., 2011; Briggs et al, 2013; Gjesteland et al., 2012; Østgaard et al., 2012]. Dwyer [2012a] has suggested a possible downward-directed positron and gamma ray signature from TGFs. Observations at ground level are necessary to observe the downward component, to better understand the TGF intensity distribution and emission pattern, to understand whether the observed 30º beaming is intrinsic to the emission process or is the result of atmospheric attenuation, and to measure the spectrum vs altitude relationship. As a practical consideration, it has been suggested that lightning-induced gamma rays might produce a significant radiation exposure for airplane passengers flying close to a lightning stroke [Dwyer et al., 2010].

The majority of ground-level observation projects currently focus on correlating satellite-observed TGFs with lightning and measuring possible associated magnetic signatures [Cummer et al. 2011; Lu et al., 2011]. The International Center for Lightning Research and Testing (ICLRT) project, however, has reported two gamma ray bursts, one in association with triggered lightning of negative polarity [Dwyer et al., 2004] and another in association with nearby negative polarity cloud-to-ground (-CG) lightning [Dwyer et al., 2012c]. TGFs associated with negative polarity lightning strikes, as with these ICLRT events, produce downward beams of photons which can be detected from the ground. ICLRT operates in a triggered mode, requiring either a triggered lightning current above 6 kA or the simultaneous trigger of two optical sensors.

The array of particle detectors at Aragats Space Environment Center (ASEC) has detected thunderstorm-associated ground enhancements above 7 MeV with timescales of microseconds and tens of minutes [Chilingarian et al., 2010, 2011]. These have been detected approximately once per year and seem to be correlated with –IC lightning.

In addition, a mountain-top detector has observed three millisecond bursts of X-rays associated with CG lightning [Moore et al., 2001]. Longer duration (40 seconds to minutes or longer) X-ray and gamma ray events have been reported previously from the ground [Tsuchiya et al., 2011, 2013], but the only other case in which a TGF-like event with millisecond emission of MeV gammas has been observed from within the atmosphere is the observation by the Airborne Detector for Energetic Lightning Emissions (ADELE) aboard an aircraft at an altitude of 14 km [Smith et al., 2011]. Here we present observations from July 2010 through February 2013 of twenty-four TGF-like events in which 50 keV - 2 MeV gamma rays are observed at ground level in shorter than 5 msec bursts associated with nearby negative polarity lightning.

**2. Detector Description:** The TGF and Energetic Thunderstorm Rooftop Array (TETRA) consists of an array of twelve 19 cm × 19 cm × 5 mm NaI(Tl) scintillators designed to detect the gamma ray emissions from nearby lightning flashes over the range 50 keV - 2 MeV. The scintillators are mounted in four detector boxes, each containing three NaI detectors viewed by individual photomultiplier tubes (PMTs). The boxes are spaced at the corners of a ~$700 \times 1300$ m$^2$ area on four high rooftops at the Baton Rouge campus of Louisiana State University (LSU) at latitude 30.41° and longitude -91.18 °. Unlike ICLRT, TETRA operates in a self-triggered mode, allowing for events to be recorded without requiring the direct detection of lightning.

Each TETRA detector box contains three NaI scintillator plates oriented at 30° from the zenith direction and separated by 120° in azimuth. Each NaI(Tl) crystal is hermetically sealed between a 6.4 mm thick glass optical window on one flat face and a 0.75 mm thick Aluminum entrance window on the other face. An ultraviolet transmitting Lucite lightguide is coupled to the glass window, and the light is viewed by an Electron Tubes 9390KB 130 mm photomultiplier tube with a standard bialkali photocathode. The scintillator-PMT assemblies are housed in ~ 1″ thick plastic foam insulation to prevent rapid temperature changes. Electronics boards in each detector box supply high voltage, amplify and shape the PMT outputs, provide an internal trigger for the data acquisition software, digitize the data, assign timestamps, and record ADC values for each event. Once triggered, each PMT anode output is integrated and assigned a 12-bit ADC value. A 32-channel 12-bit analog-to-digital converter (ADC) board, a Lassan iQ GPS board, and a Mesa FPGA board are incorporated onto a PC104 stack controlled by a Microcomputer Systems VDX-6357 800 MHz 486 CPU board running a QNX operating system. The FPGA is programmed to handle trigger logic, clock functionality, and event time stamping. We refer to the Mesa board together with its FPGA as the Trigger Logic Module (TLM). Each is capable of detecting events at a sustained rate of 30 kHz and a burst rate of up to 70 kHz. The data are then transferred over a wireless link to a central station for analysis. The initial version of the data acquisition software, used from October 2010 to January 2013, utilized a network time protocol to keep timestamps accurate to within approximately 2 msec and to monitor the absolute timing uncertainty. The current version of the software, implemented in January 2013, uses a GPS-disciplined clock to produce timestamps accurate to within 200 ns.

The ADC-to-energy conversion is calibrated with radioactive sources ($^{22}$Na, $^{137}$Cs, $^{60}$Co). Individual detector energy resolution ranges from 9 to 13.5% FWHM at 662 keV and from 5.5 to 10.8% at 1.3 MeV. The total interaction probability in the NaI scintillators is 95% at 100 keV, 82% at 500 keV, and 10% at 1 MeV (with photoelectric interaction probabilities 93%, 26%, and 0.63% respectively). In addition to the three NaI scintillators, one detector box contains a one inch diameter by one inch thick cerium-doped lanthanum bromide (LaBr$_3$:Ce) scintillator that provides high energy resolution measurements (3.5% FWHM at 662 keV) of intense events. Beginning in October 2012, all boxes contain a bare PMT to check for electronic noise.

Data are accumulated for a day at a time for each of the four detector boxes individually. The daily analysis software selects events with signals corresponding to at least 50 keV deposited energy within 1 μsec. The data are then binned into 2 msec bins and assigned a timestamp. TETRA triggers are selected with counts/2 msec at least 20 standard deviations above the mean for the day. Once days with excessive electronic noise or other instrumental problems are removed, there are 835.09 days of live time and 1303 TETRA triggers.

**3. Results:** In Fig. 1, the heavy black line shows a time history of the count rates for the three NaI photomultiplier tubes of > 50 keV events in a single detector box for one day. The total count rate, plotted in counts per minute, is reasonably constant for the first seventeen hours, and then increases by a factor of approximately 2 beginning at about 1800 CST. The small peak in the count rate seen at about 1200 CST is due to noise in the system seen only in a single PMT on a 60-second timescale. The thin black histogram near the bottom shows the local radar reflectivity in decibels acquired from www.wunderground.com, indicating rain, thunderstorms, hail, or strong winds. The increase in the NaI detector rate is clearly correlated with rainstorms. The gamma ray spectrum, measured during a rain event with the high resolution LaBr$_3$:Ce detector mounted together with the NaI detectors in one of the detector boxes, shows a clear indication of 295, 352, 609, 1120, and 1764 keV Bi$^{214}$ and Pb$^{214}$ lines characteristic of radon decay (Fig. 2).

The filled rectangle near the bottom of Fig. 1 at approximately 1800 CST marks the times of lightning strikes detected by the US Precision Lightning Network (USPLN) Unidata Program within 5 miles of the LSU campus. These are mainly cloud-to-ground events with positions accurate to approximately ¼ - ½ mile. In the upper section of the diagram, the line of filled circles marks 60-second intervals in which the NaI detector count rate is 3 standard deviations higher than the average rate for the day; these are correlated with the peak of the extended rise at the time of the rainstorms. TETRA triggers are defined as intervals during which the rate in a 2 msec window exceeds the day's average by 20 $\sigma$. The TETRA trigger observed is indicated near the top of the plot as an open square. (For a typical average counting rate of 8900 $min^{-1}$ in a detector box above 50 keV, a 20 $\sigma$ excess corresponds to 10 counts in the three PMTs in a detector box within a 2 msec window.)

Fig. 3 shows an expanded view of the data on the same day, illustrating the correlation of the triggers in individual boxes with lightning and cloud density overhead. Panel A shows the times of the triggers in each detector box. Panel B shows the rate per second of lightning strikes within 5 miles of the detectors, and Panel C shows the distance of all lightning strikes recorded by the USPLN network within 100 miles. Panel D shows the overhead cloud density.

From July 2010 through February 2013, TETRA has recorded a total of twenty-four events with triggers occurring within several minutes of thunderstorm activity producing at least one lightning flash within 5 miles of the detectors. Such events are classified as Event Candidates (ECs) and are listed in Table 1. In this table each event trigger time is listed, along with the number of lightning flashes detected within ±2.5 minutes and 5 miles and the cloud density above TETRA. Also listed for each EC is the time difference to the lightning stroke closest in time to the event trigger, the distance to that lightning stroke, the current, the number of gamma rays detected in the EC, and the $T_{90}$ duration of the event (i.e., the time over which a burst emits from 5% to 95% of its total measured counts in a single detector box). The number of sigma above the mean is listed in the second to last column for each event. (For the first three events in the table, observed simultaneously in multiple detector boxes, the smallest number of sigma above the mean is listed. These coincident events, labelled Coincident Event Candidates – CECs – are discussed in more detail below.)

TETRA's events, with an average of 20 ± 2 photons detected, are significantly smaller than the typical events observed in space. For TETRA's events, the $T_{90}$ duration was calculated by considering all events detected within a ±3 msec window around the trigger time, discarding the first and last 5% of timestamps for each event, and recording the time difference between the first and last events remaining. The uncertainty in the $T_{90}$ determination is approximately ±200 μsec based on Monte Carlo simulations of the data.

In each of the 24 events, 7 to 45 γ-rays were detected within a time window of less than 5 msec, with the total energy deposited per event ranging from 2 to 32 MeV. The distances to the nearest lightning flashes were 0.4 - 2.9 miles. For 14 events, absolute timing was available with ~2 msec accuracy. For each of these 14 events, lightning was observed within 7 seconds of the trigger time. Nine of these events were associated with -CG lightning detected within 6 msec of the trigger. Another 10 ECs were detected during June - July 2012 during a period when accurate trigger-lightning time differences were not recorded due to network timing difficulties. Eight of the ECs during that period were correlated with two intense thunderstorms that passed directly over TETRA on 6/6/2012.

The accidental rate of triggers coincident within 7 sec of a lightning flash that is less than 5 miles distant (i.e., events masquerading as ECs) is calculated based on the rate of TETRA triggers (due mainly to cosmic ray showers), the live time, and the duration of storm activity. The storm activity time is taken to be the sum of all time windows where there was lightning within 5 miles and 7 seconds and there was no electronic noise or other instrumental problems. For a total storm time of 12.65 hrs, we calculate the expected number of ECs due to accidental triggers to be 0.82. This assumes 100% lightning detection efficiency. The efficiency of the USPLN in our area has not been tested, however if we assume a similar sensitivity to that measured by Jacques et al. (2011) for cloud-to-ground lightning with peak current in excess of 20 kA of approximately 25% to account for undetected lightning flashes, then we would expect 3.3 accidental ECs compared to the 14 observed.

Fig. 4 compares data acquired within 7 seconds of lightning to the remaining data with accurate timing information. The distribution of events vs σ within 7 seconds of a USPLN lightning strike within 5 miles is shown in black. The significance distribution of the remaining data has been normalized to the total storm activity time of the lightning distribution for comparison, shown in grey. The excess of events above 20 sigma in the lightning distribution (black) as compared to the normalized distribution (grey) indicates the association of the gamma ray events with nearby lightning. (Note that, since three events involve seven separate coincident triggers in individual detector boxes, there are 18 individual triggers shown in Fig. 4 compared to the 14 ECs with accurate timing information in Table 1.) A Kolmogorov-Smirnov test of the two distributions results in a D parameter of 0.25, corresponding to high confidence that the two distributions are distinct.

The dark solid line in Fig. 5 shows the deposited energy spectrum of the 24 Event Candidates, with events observed up to 2.7 MeV deposited energy. It should be emphasized that, with TETRA's thin detectors, only a portion of the incident gamma ray energy is actually detected. Between 200 keV and 1.2 MeV, the EC spectrum is fit with a power law $E^{-\alpha}$, with $\alpha = 1.20 \pm 0.13$ and $\chi^2$/degree of freedom = 0.4 (dark dashed line). On the same figure, the grey line shows the spectrum of non-EC triggers (i.e., triggers not associated with lightning within 5 miles and 7 seconds ); this spectrum is softer, with a best fit power law index $\alpha = 1.79 \pm 0.04$ and $\chi^2$/degree of freedom = 1.0 (grey dashed line). The associations of the events reported here with negative polarity lightning strikes and the low likelihood that these are background events, along with the durations observed, are indicative of downward directed TGFs produced by the RREA mechanism.

In three of the 24 ECs, triggers were recorded in two or more boxes separated by ~1000 m within less than ±2 msec. This is approximately the relative timing accuracy between separate boxes. All three of these Coincident Event Candidates (CECs) occurred in July and August of 2011, when storms in southern Louisiana tend to be associated with disturbances in the Gulf of Mexico rather than frontal lines. No CECs were detected when there was no lightning activity within 5 miles.

Time histories for the three CECs are shown in Fig. 6. The plot shows a 50 msec window centered on the event trigger time, defined as the center of the first 2 msec bin containing a trigger. The counts for each box (i.e., the number of phototubes detecting a signal with amplitude in excess of 50 keV within the 1 microsecond PMT anode output integration time) are plotted vs time relative to the event trigger time. For the two events on 7/31/2011 (panels A and B), the lightning strikes closest in time occurred within approximately 6 and 4 msec of the event trigger. For those cases, the time of the lightning strike is shown as an X with a timing uncertainty of ±2 msec near the top of the plot. In the first 7/31/2011 event (Panel A of Fig. 6), one PMT in box #3 fired, followed by two PMTs in box #4 2.3 msec later. The distance between the two boxes was 1500 m, corresponding to a gamma ray travel time difference of up to 5 μsec. In fact, we believe the differences between the event times in the separate boxes in Fig. 6 are a direct measure of the absolute timing differences between the boxes.

The expected number of CECs due to random triggers is small: Given an initial EC with counting rate in one box in excess of 20 σ above the daily average, the likelihood that a second or third trigger occurred at random in another box within the timing uncertainty of 2 msec on the same day is estimated as (4 msec × N/86400 sec)$^{b-1}$, where N is the total number of random 20 σ triggers detected per day through February 2013 and b is the number of boxes triggered in the event. (For simplicity, we neglect here the increase in trigger rate during a thunderstorm shown in Fig. 1.) Multiplying by the number of ECs then gives the expected number of spurious CECs involving two boxes occurring by chance as 1.7 x 10$^{-6}$, as listed in Table 1.

A composite energy spectrum summed over the 3 CECs is shown in Fig. 7. A total of 80 gamma ray pulse heights above 50 keV and within the T$_{90}$ interval of each coincidence trigger are shown. The average photon energy detected is approximately 0.5 MeV, an energy at which the fraction that passes through a nominal 1 mile of atmosphere at ground level (STP) without interaction is ~ 10$^{-7}$. This average energy is low compared to the typical energies observed by the orbiting detectors (Dwyer et al., 2012b) and is presumably biased to low energies by the 0.5 cm thickness of the TETRA NaI scintillators.

Fig. 8 shows the distance from the detectors and the measured current for each lightning flash within 5 miles of TETRA from 7/1/2010 to 2/28/2013. There was a total of 5360 flashes within 5 miles. For each of the 10 ECs and CECs with lightning within 5 miles and ±100 msec of the trigger time, the distance and measured current are plotted with black X's. Although all the TETRA events correspond to lightning less than 3 miles away, the two lightning flashes within ±100 msec of a CEC are both more than a mile away. No CECs were detected with closer events. If all discharges produce TGFs (Østgaard et al., 2012), then the rate of detection and the CEC distances point to either a range of intensities extending below the sensitivity limit of TETRA, strongly non-isotropic emission, or the possibility that the gamma ray emission is only indirectly associated with the lightning (Connaughton et al., 2013). This can also occur if some gamma ray events are produced by intracloud (IC) strikes, since the USPLN data record primarily cloud-to-ground strikes.

Out of the 10 ECs shown in Fig. 7, nine were found to be within 6 msec of a negative polarity CG lightning strike within 3 miles with current above 20 kA (Table 1). For the two CECs that occurred on 7/31/2011, lightning strikes are recorded at 6 msec and 4 msec before the TETRA triggers. In both cases, these were nearby, cloud-to-ground events at 1.4 miles distance with current -43.6 kA and 1.8 miles distance with current -29.1 kA. For four ECs with accurate timing information, the lightning strikes closest in time to the TETRA triggers were in excess of ±100 msec before or after the NaI signal and so are not considered coincident with a USPLN observed strike. Again, this can occur if some gamma ray events are produced by intracloud (IC) strikes or if the gamma rays are not all directly associated with the lightning.

**4. Conclusions:** The gamma ray events described here have durations ranging from 24 μsec to 4.2 msec. The similarity of these event durations observed by TETRA to those reported by BATSE, RHESSI, AGILE, Fermi, and ICLRT suggest that the TETRA events are also generated by the RREA mechanism. Dwyer et al (2012c) compared the spectrum of x-rays from lightning to gamma rays from TGFs, showing a marked difference above 2 MeV, but the restricted energy range of TETRA and the low statistics make it impossible to draw strong conclusions from the observed TETRA spectra.

Fermi GBM data suggest that WWLLN detects shorter duration TGFs more efficiently than the longer ones because of the frequency constraints of the network (between 6 and 18 kHz). For the sferic signals found within a 400 μs window around the TGF gammas, the stronger sferics appear due to the TGF itself while the weaker sferics are due to associated +IC lightning [Connaughton et al., 2013]. The brightest TGFs seen by BATSE, RHESSI, and GBM produce ~$10^{17}$ runaway electrons with a source altitude ~ 13 km [Briggs et al., 2010]. In contrast, the two TGFs previously reported from the ground by ICLRT are associated with -CG lightning [Dwyer et al., 2004, 2012c]. The 2009 ICLRT event produced ~$10^{11}$ runaway electrons and was observed at a distance of ~2 km. If the TETRA events are characterized by typical energy 500 keV and distance 1 mile, then atmospheric absorption attenuates the flux by a factor of ~ $4 \times 10^{-8}$ at sea level.

Assuming isotropic emission at a distance of 1 mile, a typical total of 20 photons observed in an event by TETRA then requires in excess of ~$10^{18}$ photons at the source. Either the ground level TETRA events are beamed, or they are distinctly different from the ICLRT events.

Here we have presented data for a series of gamma ray events observed with a self-triggered ground array, suitable for observing weak events from nearby distances without a bias caused by a lightning trigger, and find that events with durations < 5 msec and detected individual photon energies up to at least 2 MeV appear to be produced in conjunction with nearby -CG lightning. In two CECs, these are most closely associated with -CG events 1.4 and 1.8 miles away. In the other CEC event, the nearest detected lightning strike in time is more than 6 seconds after the gamma ray event. Either this gamma ray event is not correlated with nearby lightning, the associated CG lightning strike was missed by the lightning network, or the event was due to IC lightning that was not detected by the lightning network.

**Acknowledgements:** The lightning data were provided by the US Precision Lightning Network (USPLN) Unidata Program. Radar data were downloaded from http://www.wunderground.com. The authors would like to thank J. Isbert, B. Ellison, D. Smith and N. Cannady for their assistance with the project, and especially J. Fishman for valuable discussions. This project has been supported in part by NASA/Louisiana Board of Regents Cooperative Agreement NNX07AT62A and the Curry Foundation. R. Ringuette appreciates graduate fellowship support from the Louisiana Board of Regents.

| Date | Trigger Time (CST) (hh-mm-ss) | Max Lightning Rate/sec within 5 mi. | Cloud Density (dBZ) | # Flashes within 5mi. and 5min. | Trigger-Lightning Difference (ms) | Lightning Distance (mi.) | Lightning Current (kA) | $T_{90}$ Duration (us) | Total γ rays Detected | Total Energy (MeV) | σ Above Mean | Prob. of CEC |
|---|---|---|---|---|---|---|---|---|---|---|---|---|
| 7/31/2011 | 16-21-44.976 | 2 | 45 | 12 | -6 | 1.4 | -43.6 | 702 | 22 | 14.7 | 25.2 | 1.7E-06 |
| 7/31/2011 | 16-21-45.300 | 2 | 45 | 12 | -4 | 1.8 | -29.1 | 1326 | 24 | 11.7 | 25.2 | 1.7E-06 |
| 8/18/2011 | 17-57-38.984 | 4 | 50 | 40 | 6743 | 1.3 | -23.4 | 1318 | 40 | 20.3 | 22.5 | 1.2E-13 |
| 2/24/2011 | 23-11-15.787 | 3 | 45 | 1 | -6 | 2.9 | -20.9 | 953 | 20 | 1.7 | 24.6 | - |
| 7/29/2011 | 10-38-58.932 | 6 | 45 | 42 | 5 | 0.4 | -57.7 | 153 | 8 | 4.8 | 23.0 | - |
| 8/18/2011 | 17-57-39.202 | 4 | 50 | 40 | 6525 | 1.3 | -23.4 | 24 | 7 | 3.6 | 26.1 | - |
| 3/12/2012 | 11-30-16.500 | 6 | 45 | 4 | 5 | 1.6 | -81.3 | 1997 | 7 | 3.2 | 21.8 | - |
| 4/2/2012 | 12-29-30.554 | 3 | 50 | 8 | 6 | 0.6 | -29.9 | 464 | 30 | 31.6 | 104.3 | - |
| 4/4/2012 | 02-49-21.900 | 5 | 55 | 21 | -3 | 1.9 | -158.4 | 515 | 24 | 21.3 | 88.6 | - |
| 8/5/2012 | 14-43-35.661 | 7 | 40 | 16 | -849 | 0.6 | -56.5 | 392 | 18 | 12.4 | 40.6 | - |
| 8/6/2012 | 19-17-33.359 | 5 | 50 | 1 | 1017 | 0.8 | -23.1 | 465 | 13 | 4.5 | 21.9 | - |
| 8/9/2012 | 15-27-29.804 | 4 | 50 | 21 | 2 | 0.4 | -27.8 | 2412 | 12 | 2.9 | 29.0 | - |
| 8/9/2012 | 15-28-36.070 | 4 | 50 | 27 | 80 | 0.9 | -36.7 | 4217 | 24 | 7.4 | 41.3 | - |
| 8/9/2012 | 15-28-36.560 | 4 | 50 | 27 | 2 | 0.8 | -19.2 | 146 | 12 | 8.0 | 33.9 | - |
| 6/6/2012 | 15-44-18 | 6 | 55 | 16 | - | - | - | 609 | 14 | 8.5 | 45.7 | - |
| 6/6/2012 | 15-37-31 | 6 | 55 | 40 | - | - | - | 865 | 45 | 27.2 | 86.1 | |
| 6/6/2012 | 19-23-27 | 6 | 55 | 40 | - | - | - | 2979 | 18 | 6.7 | 45.7 | |
| 6/6/2012 | 19-29-43 | 6 | 55 | 33 | - | - | - | 2376 | 24 | 9.7 | 55.3 | |
| 6/6/2012 | 19-31-21 | 6 | 55 | 19 | - | - | - | 919 | 40 | 29.8 | 48.3 | |
| 6/6/2012 | 19-32-41 | 6 | 55 | 19 | - | - | - | 827 | 9 | 5.4 | 21.0 | |
| 6/6/2012 | 19-36-40 | 6 | 55 | 18 | - | - | - | 2035 | 8 | 5.2 | 20.1 | |
| 6/6/2012 | 19-36-41 | 6 | 55 | 18 | - | - | - | 631 | 32 | 31.4 | 63.1 | |
| 6/9/2012 | 13-40-16 | 6 | 50 | 1 | - | - | - | 1930 | 15 | 8.6 | 55.1 | |
| 7/7/2012 | 17-38-45 | 5 | 45 | 1 | - | - | - | 510 | 14 | 8.0 | 33.9 | - |

**Table 1:** Properties of the 24 Event Candidates. CECs are listed in the top section; ECs for which the absolute timing uncertainty is known are listed in the middle section; and ECs for which the absolute timing uncertainty is unknown are listed in the bottom section of the table. The date and time of each EC trigger are listed, along with the properties of the storm associated with each event. The properties of the associated lightning, event duration, number of gamma rays detected, total energy and event significance are also listed for each event. The probability of each CEC occurring is listed in the last column for the CECs.

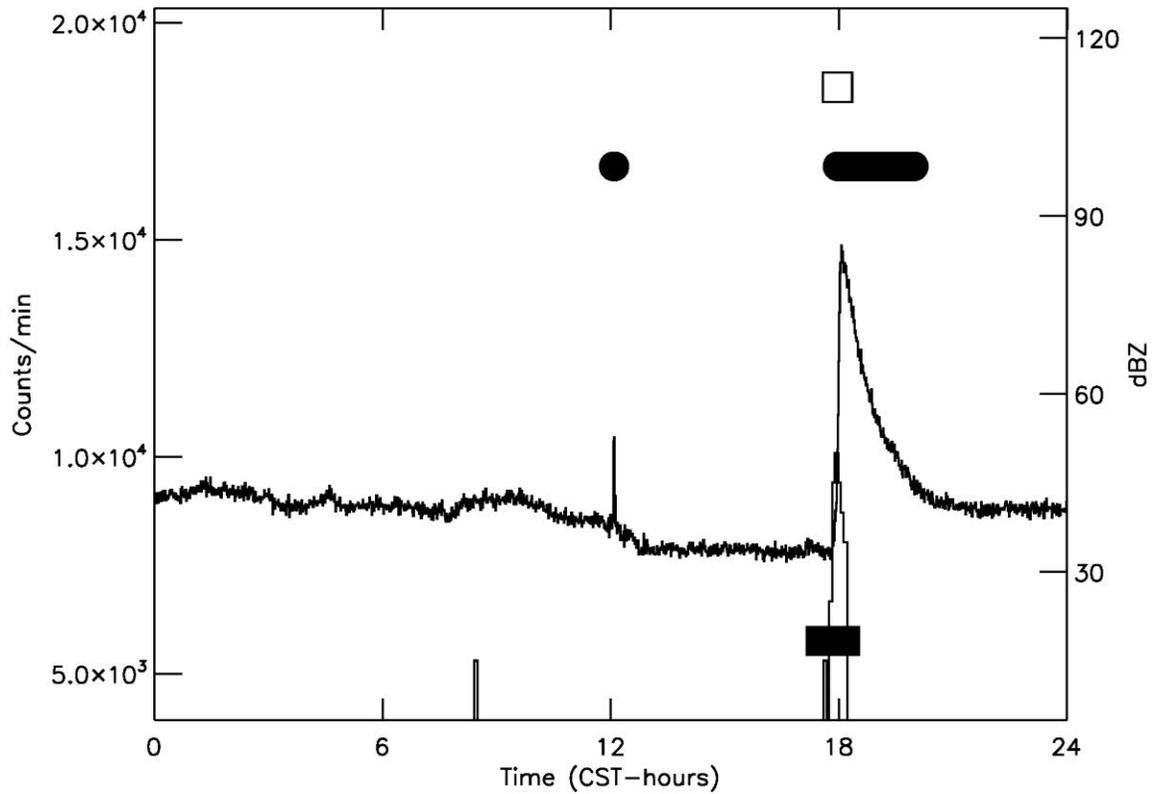

**Fig. 1.** Summed NaI counting rate per minute in Box 3 on 8/18/2011 (heavy black line, left hand scale). Thin black histogram near the bottom (right hand scale) shows radar reflectivity. The filled rectangle at the bottom marks times of lightning strikes within 5 miles. The row of filled circles near the top marks intervals in which the count rate in 60 sec bins exceeds the day's average by 3 σ; the open square marks the TETRA trigger, i.e., the interval when the rate in 2 msec bins exceeds the day's average by 20 σ.

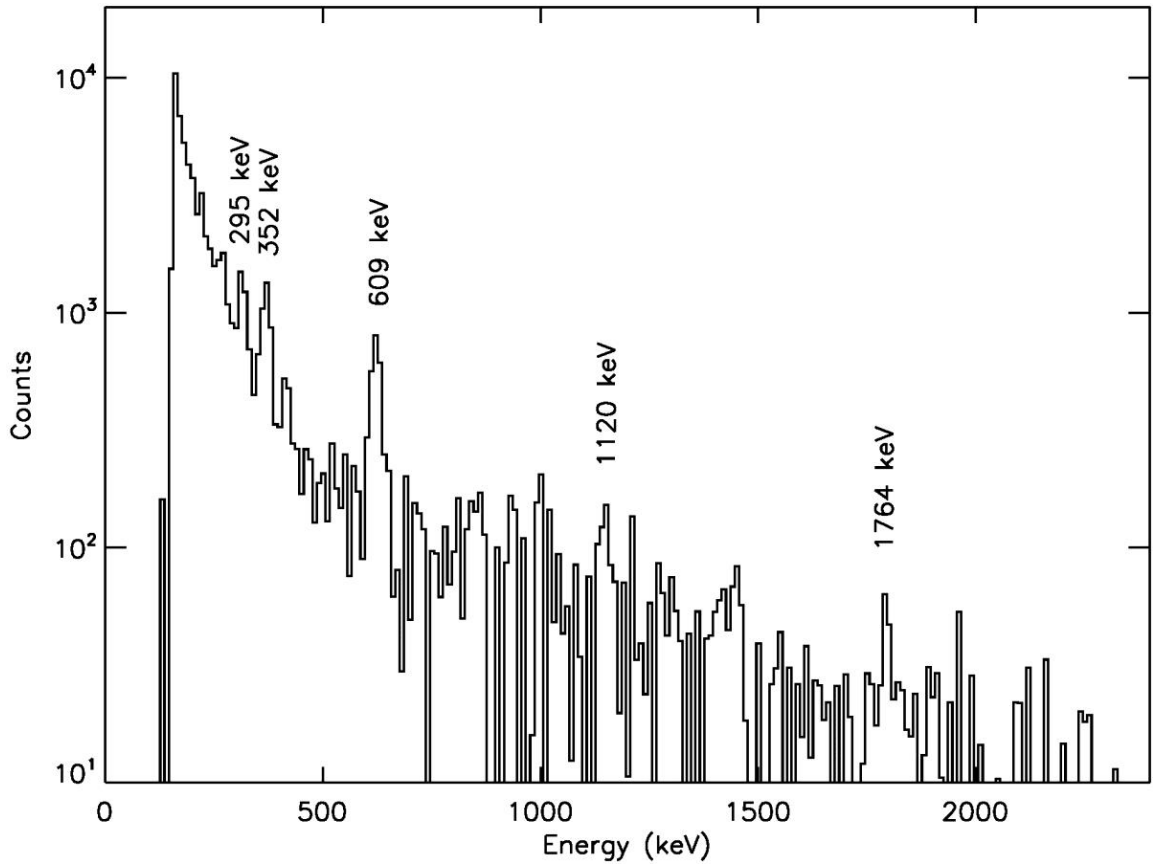

**Fig. 2.** LaBr$_3$:Ce Rain Spectrum. LaBr$_3$:Ce background-subtracted spectrum during a 6 hour precipitation event showing radon lines at 295 keV, 352 keV, 609 keV, 1120 keV and 1764 keV.

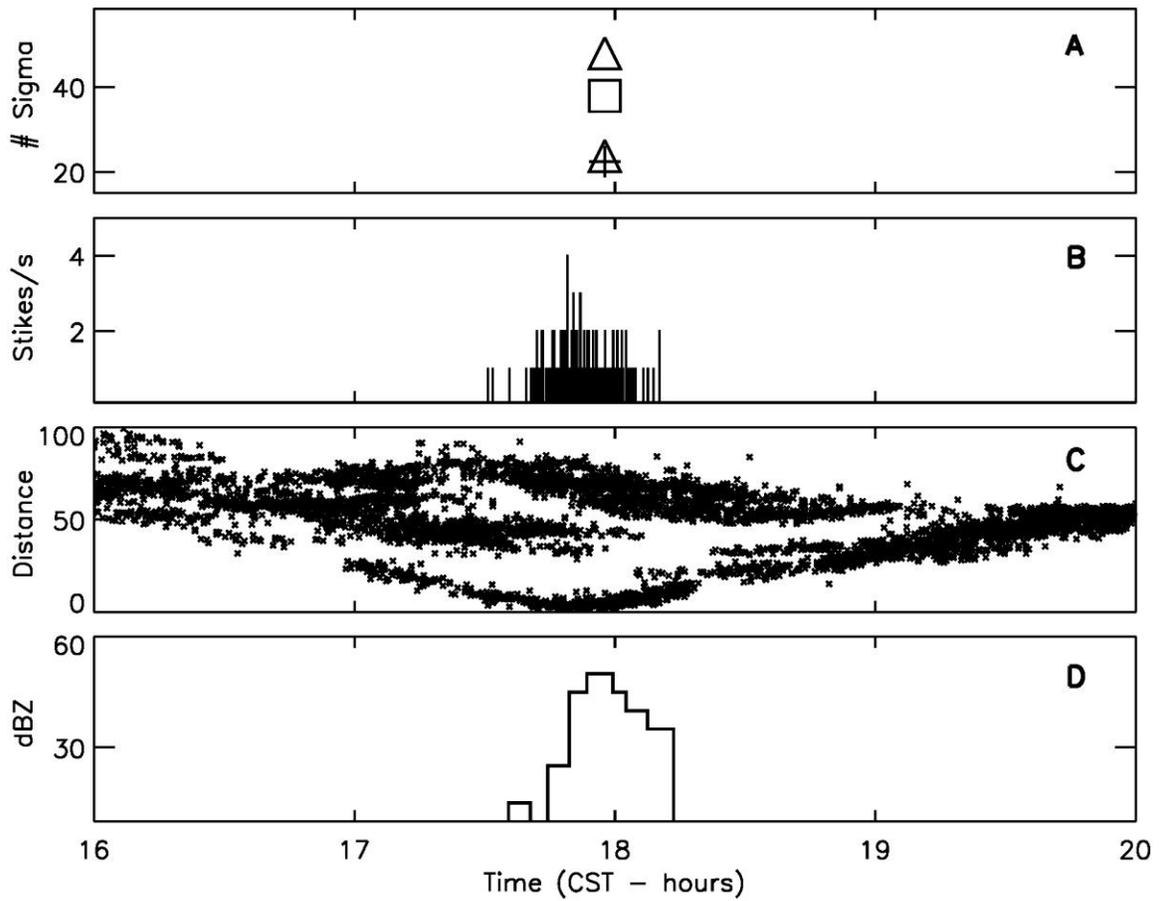

**Fig. 3.** TETRA Report for 8/18/2011 events. *Panel A (top)*: Triggers detected on 8/18/2011 (NaI signals above 50 keV in a single detector box with count rate per 2 msec in excess of 20 σ above the 8/18/2011 daily mean counting rate). Box 1 triggers are indicated by plus signs, Box 3 by triangles and Box 4 by squares. *Panel B*: Rate per second of USPLN lightning strikes within 5 miles. *Panel C*: Distance to each recorded lightning strike within 100 miles. *Panel D*: Overhead cloud density.

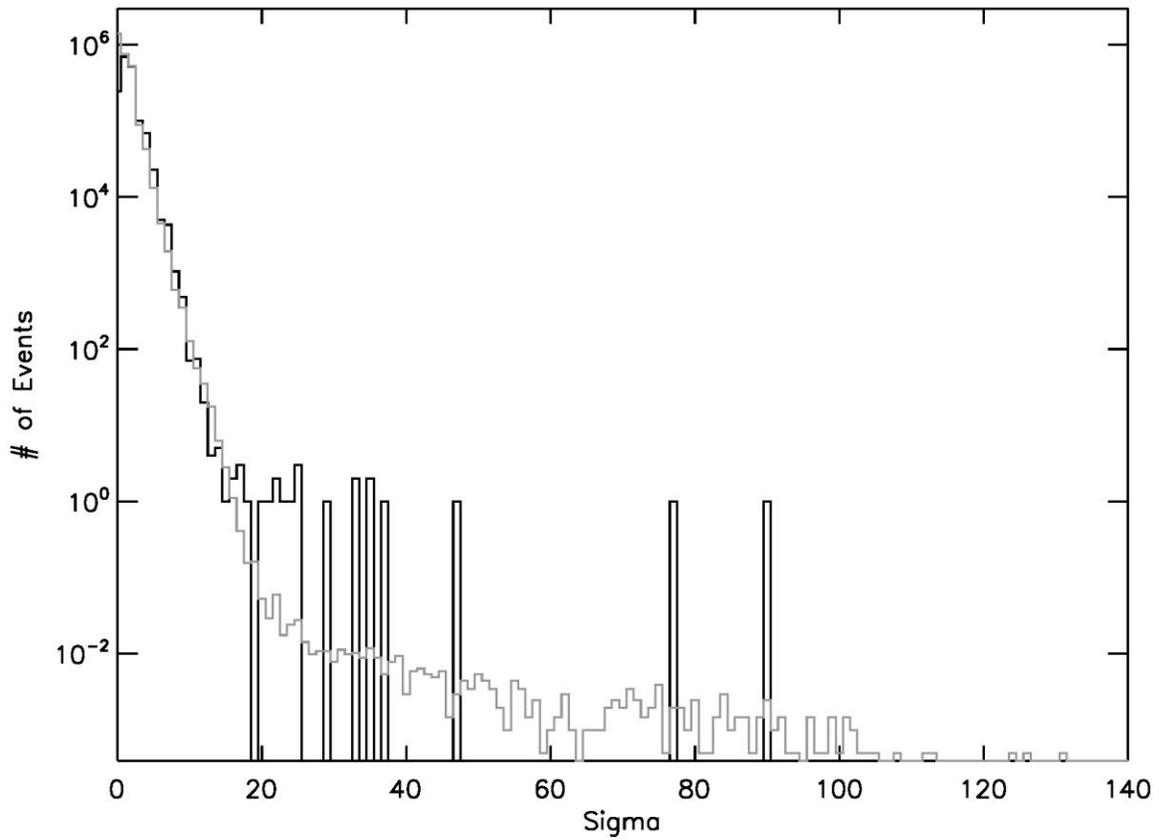

**Fig. 4.** Distribution of events with significance σ. Distribution of events within 7 seconds of nearby (< 5 miles) lightning is shown in black. Distribution of all data, normalized to 0.52 days of live time, is shown in grey, showing excess of lightning-associated ECs at σ > 20.

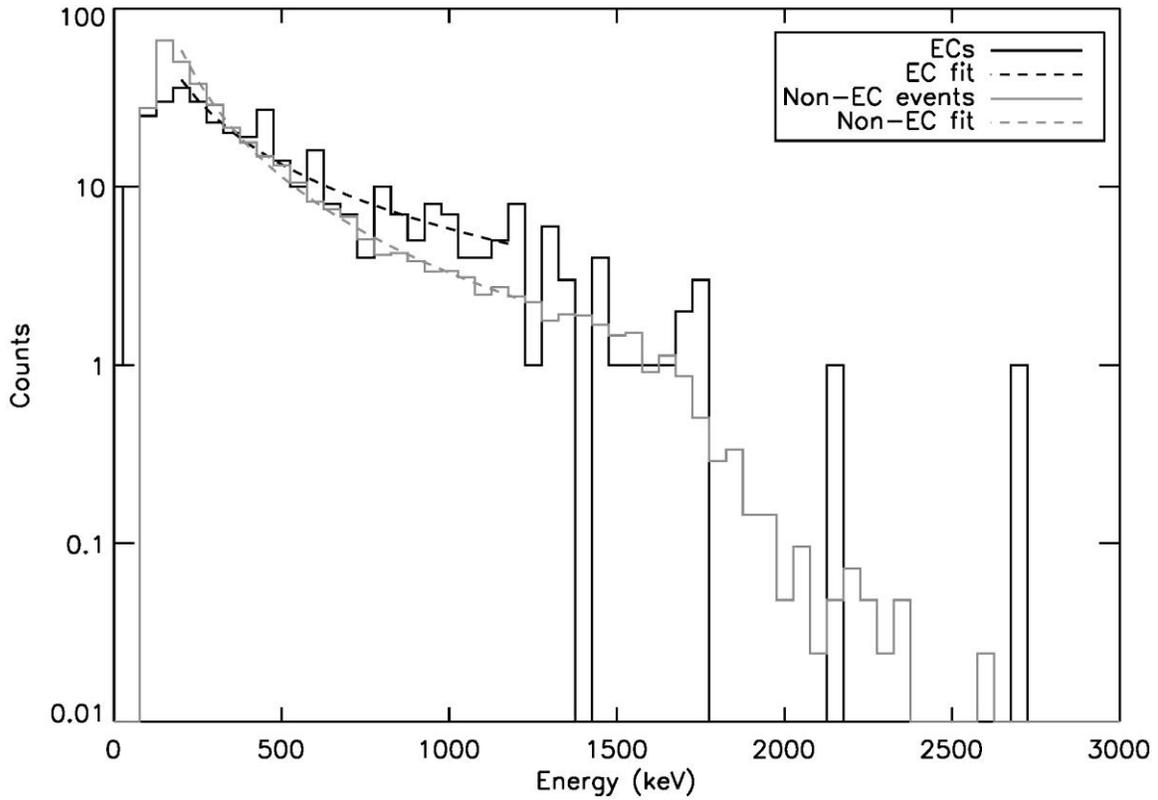

**Fig. 5**. Spectra of Event Candidates and non-EC TETRA triggers. Spectrum of ECs is shown in black. Spectrum of non-EC TETRA triggers (triggers not associated with lightning nearby in time and distance) is shown in grey. Power law fits between 200 keV and 1200 keV of the form $E^{-\alpha}$ are shown with dotted lines, where $\alpha = 1.20 \pm 0.13$ and $1.79 \pm 0.04$ for EC and non-EC events respectively.

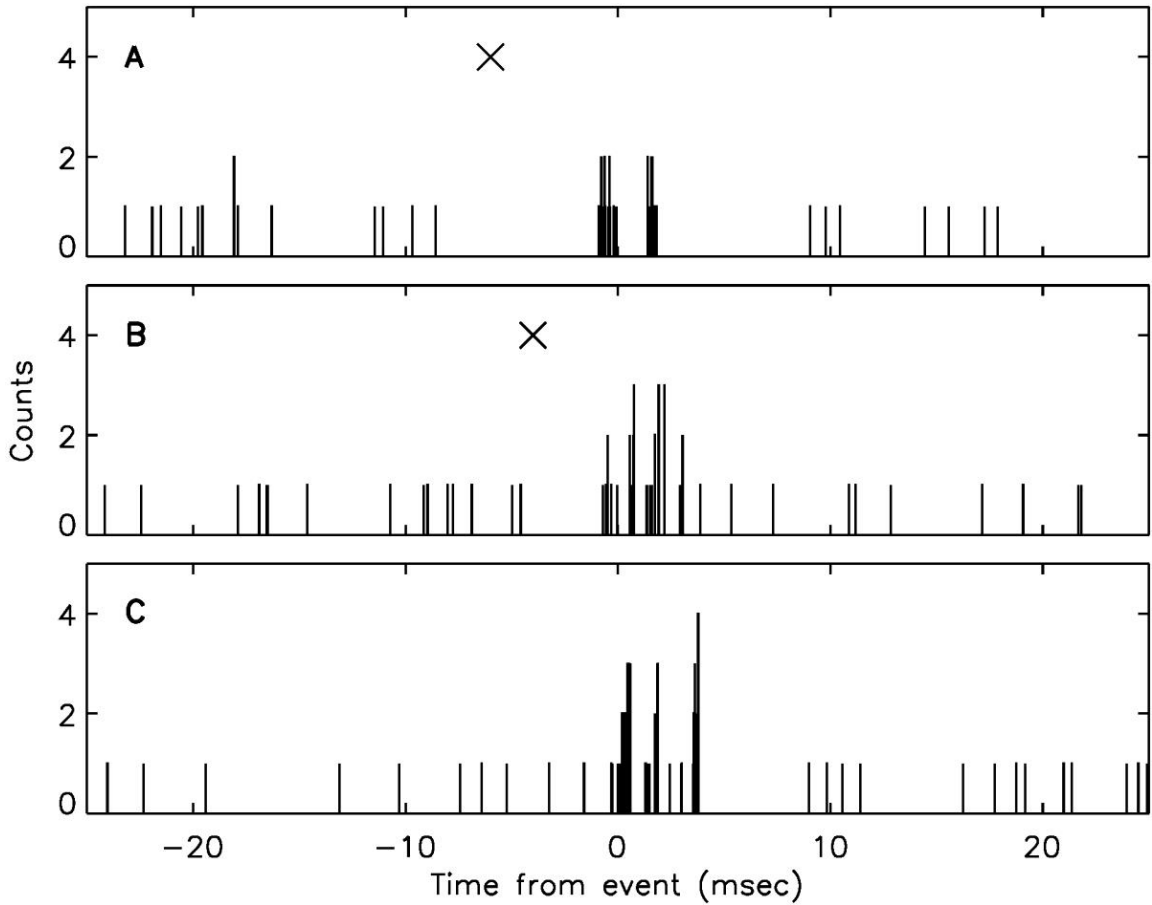

**Fig. 6.** NaI time histories over 50 msec window centered on the trigger time for each CEC. Lightning strikes within 5 miles in the 50 msec window have a ±2 msec timing uncertainty and are shown with X's in panels A and B. Panel A shows the CEC event on 7/31/2011 at 16:21:44.976 CST. The Box 3 time history is centered at 0 msec and Box 4 at 2 msec. A USPLN lightning strike within 5 miles is indicated by the X at -6 msec. Panel B shows the event on 7/31/2011 at 16:21:45.300 CST. The Box 3 time history is centered at 0 msec and Box 4 at 2 msec. A USPLN lightning strike within 5 miles is indicated by the X at -4 msec. Panel C shows the event on 8/18/2011 at 17:57:38.984 CST with the Box 3 time history centered at 0 msec, Box 1 at 2 msec and Box 4 at 4 msec. No USPLN lightning was detected within 5 miles within this 50 msec window.

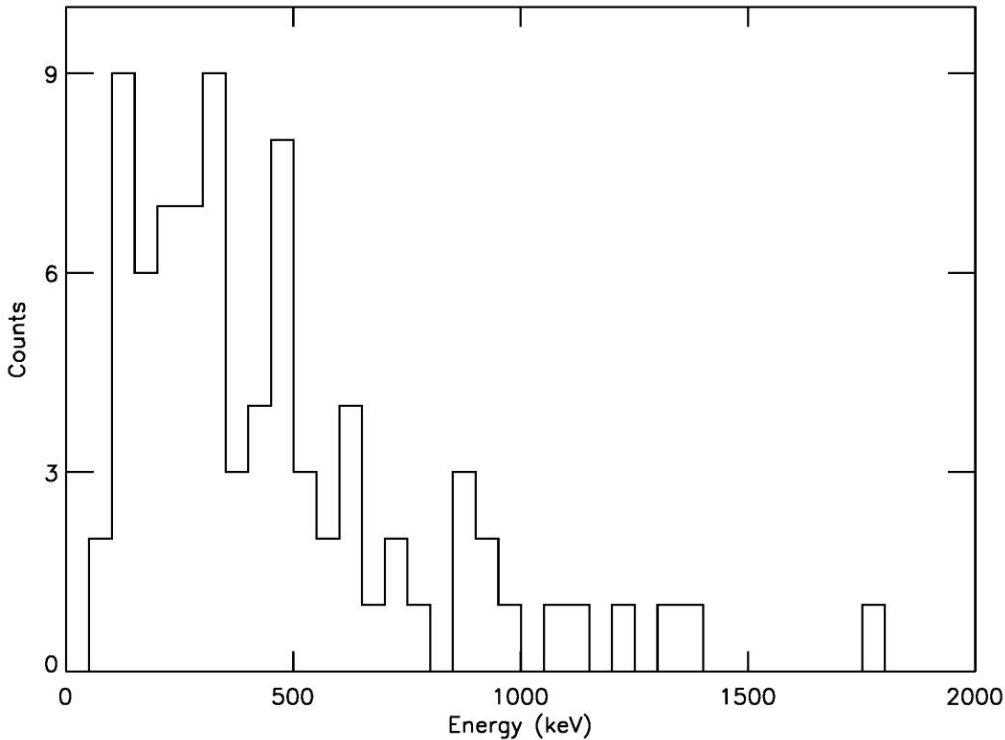

**Fig. 7.** CEC Event Spectra. Combined NaI detector energy spectrum for the three CECs. 80 photons were detected within the $T_{90}$ interval of each individual detector box's trigger time.

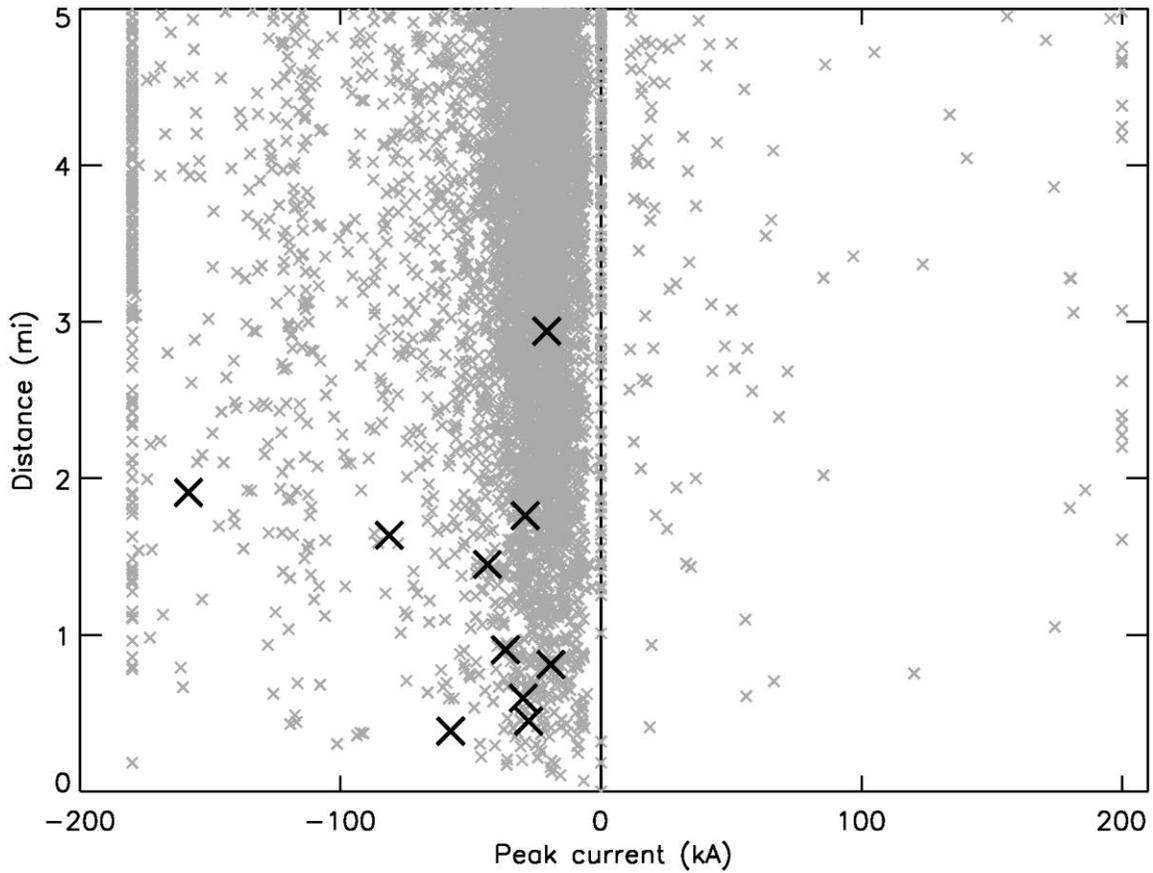

**Fig. 8.** All Lightning Activity within 5 miles of TETRA from 7/1/2010 to 2/28/2013. The current and distance for all USPLN lightning flashes within 5 miles of TETRA are indicated by grey X's. Lightning strikes that are within 5 miles and 100 msec of an EC or CEC are considered coincident strikes and are plotted with black X's. The vertical line at 0 kA indicates IC lightning.